\documentclass[fontsize=11pt,a4paper,toc=flat]{arxiv-article}
\bibliography{References,not-arxiv}

\usepackage{Definitions}

\usepackage{comment}

\usepackage{hyperref}
\graphicspath{{./Plots/}}

\usepackage[T1]{fontenc}
\usepackage[utf8]{inputenc} 
\usepackage{graphicx} 
 \usepackage{slashed}
 \usepackage{multirow}
\usepackage{slashed}
\usepackage{array}
\usepackage{booktabs}
\usepackage{amsmath}

\def\be{\begin{equation}}
\def\ee{\end{equation}}

\newcommand{\identity}{\mathbbm{1}}

\DeclareOldFontCommand{\bf}{\normalfont\bfseries}{\mathbf}
\DeclareOldFontCommand{\rm}{\normalfont}{\mathrm}

\newcommand{\OfficialTitle}{
	Resurgence Analysis of the Nambu--Jona--Lasinio model at large charge
}
\title{\setstretch{1.4}
	{\color{Thoughtless}\textls[-20]{\OfficialTitle}}
}

\hypersetup{pdfauthor={Jahmall Bersini, Simeon Hellerman, Domenico Orlando, Susanne Reffert},pdftitle={\OfficialTitle},%
	colorlinks=true,linkcolor=ThoughtYouWere,citecolor=ThoughtYouWere,urlcolor=ThoughtYouWere}

\author{%
	\begin{minipage}{.96\textwidth}
		\begin{center} \dosserif%
			{\small
				\textbf{Jahmall Bersini}\textsuperscript{\ding{74}},
				\textbf{Simeon Hellerman}\textsuperscript{\ding{74}},
				\textbf{Domenico Orlando}\textsuperscript{\ding{72}\ding{73}\ding{74}}, and
  				\textbf{Susanne Reffert}\textsuperscript{\ding{73}\ding{74}} 
			}
		\end{center}
		\authorBlock{\ding{74}}{\dosserif{}%
        Kavli Institute for the Physics and Mathematics of the Universe (WPI)\\ UTIAS, The University of Tokyo\\ Kashiwa, Chiba 277--8582, Japan}
  		\authorBlock{\ding{72}}{\dosserif{}%
			INFN sezione di Torino.\\
			via Pietro Giuria 1, 10125 Torino, Italy}
      \authorBlock{\ding{73}}{\dosserif{}%
			Albert Einstein Center for Fundamental Physics,\\
			Institute for Theoretical Physics, University of Bern,\\
			Sidlerstrasse 5, CH-3012 Bern, Switzerland}
	\end{minipage}
}

\date{}

\begin{document}

\numberwithin{equation}{section}

\begin{titlepage}

	\maketitle

	\thispagestyle{empty}

	\vfill\dosserif{}

	\abstract{ \normalfont{}\noindent{}%
     We study the fixed point of the three-dimensional \acs{njl} model in a double-scaling limit where both the charge $Q$ and the number of fermion flavors $N$ become large with a fixed ratio $q=Q/(2N)$.
     While a similar analysis has been performed for the bosonic O(N) model, fermionic models pose new challenges.
     In this work, we systematically explore the \acs{cft} spectrum in both the large and small $q$ limits beyond the first few orders, and perform a resurgence analysis.
     Through this approach, we identify the exponential corrections that relate the convergent small-\(q\) expansion to the asymptotic large-\(q\) behavior.
     Our results are suggestive of a geometric interpretation in terms of the worldline of particles moving along the geodesics on the cylinder.
}
	\vfill

\end{titlepage}

\setstretch{1.1}
\tableofcontents

\newpage

\section{Introduction}

The study of \acp{qft} at large conserved charge has emerged as a powerful approach for understanding strongly coupled dynamics~\cite{Hellerman:2015nra,Gaume:2020bmp}. By exploiting the state-operator correspondence, one can map the problem of computing the conformal dimensions of charged operators to an analysis of the system’s ground-state energy in a fixed-charge sector:
\begin{equation}
\Delta(Q) = r F(Q).
\end{equation}
Combining large charge with a second controlling parameter is often very fruitful, as was done in~\cite{Alvarez-Gaume:2019biu,Giombi:2020enj,Dondi:2024vua}, where a double-scaling limit $Q\to \infty,\ N\to \infty, \ q=Q/(2N)$ fixed was taken in the $O(N)$ vector model, or in~\cite{Badel:2019oxl,Arias-Tamargo:2019xld,Badel:2019khk,Antipin:2020abu,Sharon:2020mjs, Antipin:2022naw,Antipin:2023tar,Watanabe:2019pdh}, where a similar approach was followed combining the $\varepsilon$ expansion with large charge.
Also in the study of fermionic models, such a natural double-scaling limit exists, where both the charge \(Q\)  and the number of fermion flavors \(N\) become large while their ratio remains fixed.
The \ac{njl} model in three dimensions provides an ideal setting for exploring this approach, as it exhibits a well-defined \ac{uv} fixed point in the large-charge limit. The \ac{njl} model has already been studied in this double-scaling limit in~\cite{Dondi:2022zna}. 
The goal of this paper is to analyze the \ac{njl} model in the large and small $q$ limits systematically beyond the first few orders, establishing a controlled expansion for its physical observables. We moreover perform a resurgence analysis, identifying the exponential corrections in analogy with~\cite{Grassi:2019txd,Hellerman:2018xpi,Hellerman:2021yqz,Hellerman:2021duh,Dondi:2021buw,Antipin:2022dsm}. 
\bigskip
 
In the regime \(q \ll 1 \), using an explicit expansion of the gap equation and grand potential \(\Omega(\mu)\), we derive a power series in $q$ for the scaling dimensions and analyze its convergence properties.
This perturbative expansion is \emph{convergent} within a finite radius.
This is a striking feature, as many \acp{qft} at large charge exhibit only asymptotic expansions in perturbation theory.
From a more technical point of view, we observe that the analytic structure of the small double-scaled coupling expansion in the grandcanonical picture is considerably more complicated than in the canonical one. 
With a careful study of the analytic properties of the perturbative expansion, confirmed by Padé approximants and Darboux’s theorem, we identify the singularity that determines the radius of convergence, which appears to be associated with a vanishing mass of the radial mode.
This observation suggests that the small-charge expansion is under analytic control, making it a valuable tool for extracting physical observables.

\bigskip

In contrast, the expansion at large double-scaled coupling \(q \gg 1\) is \emph{asymptotic}, with coefficients exhibiting factorial growth characteristic of a divergent series.
A key role is played by the gap equation, governing the large-\(q\) limit:
\begin{equation}
\kappa_0 \tanh(\kappa_0) = 1,
\end{equation}
which determines the nonzero fermion gap $\Phi_0$ in the double-scaling limit as a function of the chemical potential $\mu$ 
\begin{equation}
	\Phi_0 = \mu \sqrt{\kappa_0^2 -1} \,,
\end{equation}
and admits a positive real solution:
\begin{equation}
\kappa_0 = 1.1996786402577338339163698486411419442614 \dots \,.
\end{equation}

Using a Mellin transform representation of the relevant functional determinants, we systematically expand the energy and operator dimensions in inverse powers of \(q\):
\begin{equation}
\frac{\Delta(Q)}{2N} = \frac{2}{3} \kappa_0^{3/2} q^{3/2} + \frac{1}{6} \kappa_0^{1/2} q^{1/2} + \dots \,.
\end{equation}
The asymptotic nature of the expansion (whose terms diverge like \((2n)!\)) implies the existence of an optimal truncation of the series at \(\order{\sqrt{q}}\) terms, and the necessity of non-perturbative corrections, which we identify as exponentially suppressed terms of the form \(\exp( - \alpha\sqrt{q})\).
These corrections indicate the presence of worldline-instanton contributions to the physics of larger double-scaled coupling in the large-charge limit.

\bigskip

To better understand these non-perturbative effects, we turn to \emph{resurgence theory} indicating the effect of \emph{worldline instantons}.
The non-perturbative terms found in the asymptotic expansion can be interpreted as contributions from worldline instantons wrapping geodesics on the two-sphere. 
The scaling dimension receives nonperturbative corrections of the form
\begin{equation}
  \frac{\Delta(Q)}{2N} \ni - i \frac{ \pqty*{\kappa_0^2 - 1}^{1/4} q^{3/4} e^{-2 \pi  \sqrt{\kappa_0^{-1} - \kappa_0^{-3}}\abs{k} \sqrt{q}}}{2 \pi \kappa_0^{9/4} \sqrt{\abs{k}}}  \pqty*{ \sqrt{\kappa_0^2 - 1} \cos*(\frac{2 \pi k}{\kappa_0^{3/2}} \sqrt{q}) +  \sin*(\frac{2 \pi k}{\kappa_0^{3/2}} \sqrt{q})  } +  \dots  \,.
\end{equation}
The worldline formalism provides a first-quantized description of the \ac{njl} model’s fermionic determinant, allowing us to derive the leading non-perturbative corrections directly from a semiclassical calculation. This interpretation unifies the perturbative and non-perturbative structures of the theory, demonstrating that the large-charge expansion is naturally controlled by geometric instanton effects.

While these techniques have been previously applied to bosonic theories, their successful implementation in a fermionic model provides strong evidence that they apply more broadly.
This raises intriguing questions about the role of large charge in other strongly coupled \acp{qft}, including gauge theories and supersymmetric models.

\paragraph{Plan of the paper.}
In Section~\ref{sec:review}, we introduce the \ac{njl} model and review its large- structure, focusing on the fixed-charge sector.
Section~\ref{sec:smallQ} presents the small-charge expansion, deriving a convergent perturbative series and analyzing its analytic properties.
In Section~\ref{sec:largeQ}, we develop the large-charge expansion, demonstrating its asymptotic nature and extracting non-perturbative corrections.
Section~\ref{sec:geomInterp} provides a geometric interpretation of these corrections using worldline instantons and semiclassical techniques.
Finally, Section~\ref{sec:conclusions} summarizes our conclusions and discusses future directions. Appendix~\ref{Eulero} contains alternative all-order expressions for the gap equation and the grand potential, while Appendix~\ref{app2} provides details on the worldline representation.

\section{The 3d NJL model at large charge}
\label{sec:review}

The Lagrangian of the \ac{njl} model reads
\be \label{NJLLag}
\mathcal{L}_{NJL} = \bar\Psi_j \Gamma^\mu \del_\mu\Psi^j - \frac{g}{N} \left[ \left( \bar\Psi_j\Psi^j\right)^2 - \left(  \bar\Psi_j \Gamma_5\Psi^j\right)^2\right] ,
\ee
where $j=1,\dots, N$ and the gamma matrices $\Gamma_\mu$ correspond to the reducible four-dimensional representation of the Clifford algebra.
This representation is introduced to generalize the concept of chirality in arbitrary dimensions.
Explicitly, we have
\begin{align}
    \Gamma_\mu = \begin{pmatrix}
        \gamma_\mu & 0 \\
        0 & -\gamma_\mu
    \end{pmatrix},
\end{align}
where $\gamma_\mu = \sigma_\mu$ (Pauli matrices) are the gamma matrices in $d=3$.
The model flows to a \ac{uv} fixed point in $2<d<4$ dimensions, making it renormalizable order by order in the $1/N$ expansion. The theory has a $U(N)$ flavor symmetry and a $U(1)_A$ chiral symmetry acting as
\begin{align}
	\Psi_i &\to e^{i \alpha \Gamma_5} \Psi_i \,, & \Gamma_5 &= \begin{pmatrix}
        0 & \identity \\
        \identity & 0
    \end{pmatrix}.
\end{align}
It has been conjectured that in $2<d<4$ a \ac{uv} completion of the \ac{njl} model is provided by the \ac{njly} model,
\begin{equation} \label{NJLYLAG}
\mathcal L _\text{NJLY}=\frac{1}{2}(\del_{\mu}\phi_1)^2+\frac{1}{2}(\del_{\mu}\phi_2)^2+\bar\psi_j\slashed\del\psi^j+g\,\bar\psi_{Rj}\bar\phi\psi_L^j+g\,\bar\psi_{Lj}\phi\psi^j_R+\frac{(4 \pi)^2\lambda}{24}\left(\bar\phi\phi\right)^2,
\end{equation}
which exhibits a weakly-coupled infrared fixed point in $d=4-\epsilon$ dimensions. 

The fixed point of the \ac{njl} model in the double-scaling limit 
\begin{align}
	N &\to \infty, & Q &\to \infty, & Q/(2N) &= q \text{ fixed} 
\end{align}
has been studied in~\cite{Dondi:2022zna}.
The conformal dimension of the lowest operator of charge \(Q\) is obtained via the state-operator correspondence which identifies it with the energy on a cylinder of radius \(r\), \(\setR \times S^2\):
\begin{equation}
  \Delta(Q) = r F(Q) .
\end{equation}
At leading order in \(1/Q\), the energy is computed as the Legendre transform of the grand potential \(\Omega(\mu)\), \emph{i.e.} the energy of the system as a function of the chemical potential \(\mu\) conjugate to the axial charge:
\begin{align}
  \Omega(\mu) &= - \lim_{\beta \to \infty} \frac{1}{\beta} \log \Tr*( e^{-\beta (H + \mu Q)}), \\
  F(Q) &= \eval*{ \mu Q + \Omega(\mu)}_{\mu = \mu(Q)}\,,   & Q = - \odv{\Omega}{\mu} \, .
\end{align}
In turn, in the large-\(N\) limit, the grand potential is identified with the value of the action at the saddle.

In the standard approach, one introduces a Hubbard--Stratonovich collective scalar field \(\Phi\) to write the action as
\begin{equation}
  \label{HSaction}
  S = \int \dd^d{x} \left [ \bar \Psi_j \left ( \Gamma_\mu \partial^\mu - \mu \Gamma_3 \Gamma_5  + \Phi P_+ + \bar \Phi P_- \right ) \Psi^j + \frac{N}{4 g} \bar \Phi \Phi \right ],
\end{equation}
where \(P_{\pm} \) is the chiral projector $P_{\pm} = (1 \pm \Gamma_5)/2$.
The collective field $\Phi$ (which is the Legendre dual to the fermion bilinear \(\bar\Psi \Psi + \bar\Psi\Gamma_5 \Psi\)), is charged under the $U(1)_A$, acting as
\begin{equation}
  \Phi \to e^{-2 i \alpha } \Phi.
\end{equation}
As shown in~\cite{Antipin:2020abu,Dondi:2022wli}, for finite values of the chemical potential, the field \(\Phi\) develops a homogeneous \ac{vev}, which, with an abuse of notation since we are at finite volume, we will identify with the fermion gap. This condensate is an example of the Cooper mechanism at work. This can be explicitly shown by applying a Pauli--Gürsey transformation~\cite{pauli1957conservation,gursey1958relation}, after which the collective field takes the explicit form of a Cooper pair $\Psi^T C_4 \Psi$, see~\cite{Dondi:2022zna,Ebert:2016ygm}.

After integrating out the fermions, which now appear only quadratically, we can evaluate the action on the zero mode \(\braket{\Phi} = \Phi_0\), that we can choose --- without loss of generality --- to be real:
\begin{equation}
  S(\mu, \Phi_0) = - \beta N \Tr \log( \Gamma_\mu \partial^\mu - \mu \Gamma_3 \Gamma_5 + \Phi_0 )  \, .
\end{equation}
The action is to be minimized with respect to \(\Phi_0\) (gap equation) and the value at the saddle is the leading-\(N\) approximation of the grand potential,
\begin{align}
  \beta \Omega(\mu) &= \eval*{ S(\mu, \Phi_0)}_{\Phi_0 = \Phi_0(\mu)}  \, , & 
  \eval*{\pdv{S(\mu, \Phi_0)}{\Phi_0}}_{\Phi_0 = \Phi_0(\mu)} &= 0 .
\end{align}

The functional determinant in \(S(\mu, \Phi_0)\) requires regularization.
It is convenient to use a zeta-function scheme by defining
\begin{equation}
  \zeta(s | \slashed{\del{}}, \mu, \Phi_0) = \Tr*( \Gamma_\mu \partial^\mu - \mu \Gamma_3 \Gamma_5 + \Phi_0 )^{-2s} .
\end{equation}
Knowing the dispersion relation for the fermions and their multiplicity on the sphere, 
\begin{align}
	\omega_\pm(\ell)&=\sqrt{\left(\frac{\ell + 1}{r} \pm \mu \right)^2+\Phi_0^2}\,, & n_\ell&=2(\ell+1),
\end{align}
we can write an explicit expression for this function:
\begin{multline}
  \zeta(s | \slashed{\del{}}, \mu, \Phi_0) =  \sum_{\ell=0}^{\infty}  n_{\ell} \pqty*{\omega_+(\ell)^{-2s} + \omega_-(\ell)^{-2s}} \\
  = 2 r^{2s} \sum_{\ell=1}^{\infty}  \ell \bqty*{\pqty*{ \pqty*{\ell +  r \mu}^2+ \pqty*{r\Phi_0}^2}^{-s} +  \pqty*{ \pqty*{\ell - r \mu}^2+ \pqty*{r\Phi_0}^2}^{-s}} \, .
\end{multline}
In this scheme, the action for the zero modes is identified with a special value of the zeta function,
\begin{equation}
  S(\mu, \Phi_0) = - \beta N \zeta(-1/2 | \slashed{\del{}}, \mu, \Phi_0) \, ,
\end{equation}
and the gap equation reduces to
\begin{equation} \label{gappo}
   \Phi_0 \zeta(1/2 | \slashed{\del{}}, \mu, \Phi_0)  = 0 .
 \end{equation}
 
All the physical observables can be evaluated perturbatively in the double-scaling limit \(Q \to \infty\), \(N \to \infty\), \(Q/(2N)\) fixed.
The first terms in the expansion for large and small values of \(Q/(2N)\) have been found in~\cite{Dondi:2021buw}.
\begin{itemize}
\item In the \emph{small charge} regime \(Q/(2N) = q \ll 1\), the gap equation gives
  \begin{equation}
    \mu - \frac{1}{2 r} = r \Phi_0^2 + \frac{\pi^2 - 8}{4} r^3 \Phi_0^4 + \dots \, ,
  \end{equation}
  the grand potential is
  \begin{equation}
   \frac{\Omega}{2N} = - \frac{ \pi^2}{8 r} (r \Phi_0)^4 + \dots 
  \end{equation}
The energy reads
  \begin{equation}
\frac{r F}{2N} = \frac{1}{2} \frac{Q}{2N} + \frac{2}{\pi^2} \pqty*{\frac{Q}{2N}}^2 + \dots = \frac{1}{2} q + \frac{2}{\pi^2} q^2 + \dots \,.
  \end{equation}
\item In the \emph{large charge} regime \(Q/(2N) = q \gg 1\) we have
  \begin{align}
    r \Phi_0 &= \sqrt{\kappa_0^2 - 1} \pqty*{ r \mu - \frac{1}{12 \kappa_0^2} \frac{1}{r \mu} + \dots}, \\
    \frac{\Omega}{2N} &= - \frac{\kappa_0^3}{3} r^2 \mu^3 + \dots ,\\
   \frac{r F}{2N} &= \frac{2}{3 \kappa_0^{3/2}} q^{3/2} + \frac{1}{6 \kappa_0^{1/2}} q^{1/2} + \dots,
  \end{align}
  where \(\kappa_0\) is the solution to the equation \(\kappa_0 \tanh(\kappa_0) = 1\).
\end{itemize}
In the following we will derive an all-order expansion for these quantities and use resurgence techniques to relate the two limits and find the optimal truncation of the large-charge expansion.

\section{Convergent expansion in the small Q/N limit}
\label{sec:smallQ}

\subsection{Perturbative expansion}
\label{sec:small-Q-N-pert-expansion}

\begin{figure}
  \centering
   \begin{tikzpicture}
        \node[anchor=south west,inner sep=0] (image) at (0,0) {\includegraphics[ width=0.45\textwidth]{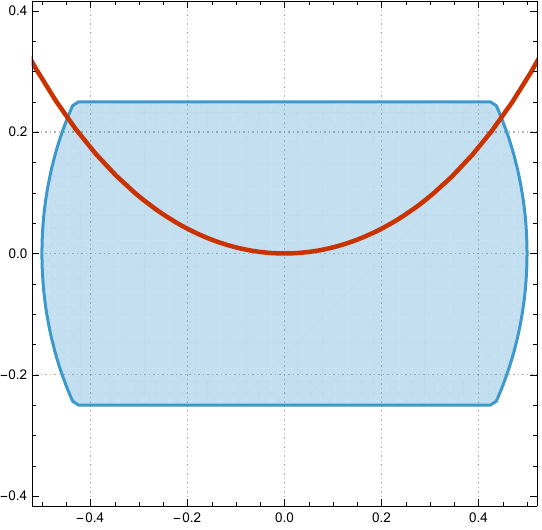}};

        \begin{scope}[x={(image.south east)},y={(image.north west)}]
          \footnotesize
            \node[] at (-0.02,1) {\(r \tilde \mu\)}; %
            \node[] at (1,0) {\(r \Phi_0\)}; %
            \node[] at (.2,.3) {\(\mathcal{C}\)}; %
        \end{scope}
      \end{tikzpicture}

	\caption{Region of convergence \(\mathcal{C}\) for the small charge expansion of the functional determinant at leading order in \(N\). The blue line represents the solution to the gap equation, intercepting the region \(\mathcal{C}\) at \(r \Phi_0 = 0.447 \dots\), \(r \tilde \mu = 0.223 \dots \). }
	\label{fig:convergence}
\end{figure}
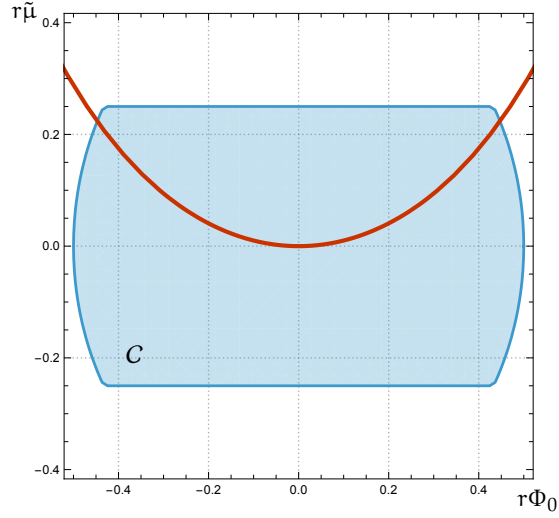

To obtain the small $q=Q/(2N)$ expansion of the gap equation and the grand potential, we first note that the conformal coupling to the Ricci scalar of $S^2$ generates a mass term $m=1/(2r)$ for the collective field $\Phi$. 
As a consequence, both the charge and the gap $\Phi_0$ vanish for $\mu = m = 1/(2 r)$.
This is the value of the chemical potential around which one has to expand to generate the small $q$ expansion.
For this reason it is convenient to separate $\mu = 1/(2r) + \tilde \mu$ and write the zeta function as
\begin{equation}
  \zeta(s | \slashed{\del{}}, \mu, \Phi_0) = r^{2s}  \sum _{\ell=1}^{\infty } \ell \bqty*{\pqty*{ \pqty*{\ell + 1/2 + r \tilde \mu}^2+ \pqty*{r\Phi_0}^2}^{-s} +  \pqty*{ \pqty*{\ell - 1/2 - r \tilde \mu}^2+ \pqty*{r\Phi_0}^2}^{-s}} \, .%
\end{equation}
In this section we are interested in the limit \(r \tilde \mu \ll 1\), \(r \Phi_0 \ll 1\), which eventually will translate into small \(q\).
In this limit we can use a trinomial expansion to write a double sum in positive powers of \(\tilde \mu\) and \(\Phi_0^2 +  \tilde \mu^2 \):
\begin{multline}
  \pqty*{\pqty*{\ell \pm 1/2 + r \tilde \mu}^2+ \pqty*{r\Phi_0}^2}^{-s} \\
  = \sum_{m,n = 0 }^{\infty}  \trinom{-s}{-s-m-n}{m}{n} r^{m + 2n} \pqty*{\ell \pm 1/2}^{2(-s-m-n)} \pqty*{2 \pqty*{\ell \pm 1/2} \tilde \mu}^{m} \pqty*{\tilde \mu^2 + \Phi_0^2}^n   \,,
\end{multline}
where
\begin{equation}
  \trinom{n}{k_1}{k_2}{k_3} = \frac{n!}{k_1! \ k_2! \  k_3!}   \ ,
\end{equation}
and express the sum over \(\ell\) in terms of Riemann zeta functions:
\begin{multline}
  \zeta(s | \slashed{\del{}}, \mu, \Phi_0)  = r^{2s} \sum_{n=0}^{\infty} \sum_{m=0}^{\infty} {\textstyle \trinom{-s}{-s-m-n}{m}{n} \frac{ 1 - 3 (-1)^m}{2}} \pqty*{2^{2 \ceil{m/2} + 2n + 2s - 1 } - 1 } \\ \zeta(2 \ceil{m/2} + 2n + 2s - 1) (2  \tilde{\mu})^m \pqty*{ \Phi_0^2 +  \tilde \mu^2 }^{n} r^{m +2 n}\ ,
\end{multline}
where
 \(\ceil{x}\) is the ceiling function (that maps \(x\) to the least integer greater than or equal to \(x\)).
From the trinomial expansion, we see that the sum is absolutely convergent in the region (see Figure~\ref{fig:convergence})
\begin{equation}
  \mathcal{C} = \set{ r \abs{ \tilde \mu} < 1/4 \cap r^2 \abs{ \Phi_0^2 + \tilde \mu^2} < 1/4} \ .
\end{equation}
A similar result, with a finite radius of convergence in the small-charge expansion, had already been found for the  bosonic \(O(N)\) model~\cite{Dondi:2021buw}.

In the cases of interest for us, \emph{viz.} \(s = \pm 1/2\), we can write the values of the zeta functions explicitly in terms of Bernoulli numbers:
\begin{align}
   \zeta(2p) = \frac{\abs{B_{2p}}(2 \pi)^{2p}}{2 (2p)!}, && p = 1, 2, \dots,
\end{align}
and we find that the action for the zero modes \( S (\tilde \mu, \Phi_0) = - \beta N \zeta(-1/2| \slashed{\del{}}, \mu, \Phi_0)\) reads
\begin{equation}
  \frac{S(\tilde \mu, \Phi_0)}{\beta N} = - \frac{\pi^2 r^2 \tilde \mu}{8} \pqty*{\Phi_0^2 +  \tilde \mu^2} - \frac{3 \pi^2 r^3 \tilde \mu^2}{8} \pqty*{\Phi_0^2 +  \tilde \mu^2} +\frac{\pi^2 r^3}{16} \pqty*{\Phi_0^2 +  \tilde \mu^2}^2 + \frac{\pi^2 r^4 \tilde \mu}{32}\pqty*{\Phi_0^2 +  \tilde \mu^2}^2 + \dots %
\end{equation}
Now we need to solve the gap equation, perform the Legendre transform and express the energy of the system as an expansion in the charge.

The gap equation is $\zeta(1/2 | \slashed{\del{}}, \mu, \Phi_0) = 0$:
\begin{equation}
  - 3 \pi^2 r^2 \tilde \mu \pqty*{\Phi_0^2 +  \tilde \mu^2} - 15 \pi^2 r^3 \tilde \mu^2 \pqty*{\Phi_0^2 +  \tilde \mu^2} + \frac{3\pi^2 r^3}{2} \pqty*{\Phi_0^2 +  \tilde \mu^2}^2 + \frac{5 \pi^2 r^4 \tilde \mu}{4} \pqty*{\Phi_0^2 +  \tilde \mu^2}^2 + \dots = 0 \ .
\end{equation}
The fact that \(\tilde{\mu}\) appears with even and odd powers while \(\Phi_0\) only appears with even powers and that for zero chemical potential \(\tilde{\mu}\) we expect no gap, suggests an ansatz of the form
\begin{equation} 
  r \tilde{\mu} = \sum_{j=1} \mu_j (r \Phi_0)^{2j} \ .
\end{equation}
Solving the gap equation order by order we find
\begin{equation} \label{gapseries}
  r \tilde{\mu} = (r \Phi_0)^2 + \frac{\pi^2 - 8}{4} (r \Phi_0)^4 + \frac{192 - 8 \pi^2 - \pi^4}{24} (r \Phi_0)^6 + \dots \ ,
\end{equation}
and substituting this expansion in the action for the zero modes, we obtain the grand potential as a function of the \ac{vev} \(\Phi_0\),
\begin{equation}
  \label{potseries}
  \frac{\Omega }{2 N}= \sum_{j=1}^{\infty} \Omega_j (r \Phi_0)^{2(j+1)} = -\frac{\pi ^2}{8} (r \Phi_0)^{4} -\frac{\pi ^4}{48} (r \Phi_0)^{6} + \dots  \ ,
\end{equation}
which is again a convergent series.
Its radius of convergence is given by the intersection of the curve \(\tilde{\mu} = \tilde{\mu}(\Phi_0)\) with the boundary of the region \(\mathcal{C}\) defined above.
Numerically we find \(\abs{r \Phi_0} < 0.447 \dots\), corresponding to \(r \tilde \mu < 0.223 \dots\) (see Figure~\ref{fig:convergence}).

\bigskip

The energy (equivalently, the conformal dimension of the lowest operator), is the Legendre transform with respect to \(\mu\),
\begin{equation}
  Q = \odv{\Omega}{\mu} = \odv{\Phi_0}{\mu}\odv{\Omega}{\Phi_{0}} = \frac{1}{\odv{\mu}/{\Phi_0}} \odv{\Omega}{\Phi_{0}}\, ,
\end{equation}
which, again, can be solved at any given order:
\begin{align}%
  \label{gapserQ}
  (r \Phi_0)^2=  \sum_{j=1}^{\infty} \phi_j \left (\frac{Q}{2N} \right)^{j} &= \frac{4}{\pi ^2}q+ \frac{4 \left(\pi ^2-16\right)}{\pi ^4}  q^{2}+ \frac{8 \left(960-96 \pi ^2+\pi ^4\right)}{3 \pi ^6} q^3 \nonumber \\ & + \frac{8 \left(-147456+19776 \pi ^2-552 \pi ^4+\pi ^6\right)}{9 \pi ^8} q^4+ \dots
\end{align}
to give the final result
\begin{equation} \label{altj}
   \frac{\Delta }{2 N} = \frac{r F}{2 N}=  \sum_{j=1}^{\infty} \delta_j \left (\frac{Q}{2N} \right)^{j} = \frac{1}{2} q + \frac{2}{ \pi ^2} q^2 + \frac{8 \left(\pi ^2-12\right)}{3 \pi ^4} q^3 +\frac{2 \left(\pi ^2-48\right) \left(3 \pi ^2-32\right)}{3 \pi ^6} q^4 + \dots 
 \end{equation}
 
One can estimate the radius of convergence for this series starting from what we had found for \(\Omega(\mu)\).
The computation is numerically challenging because it is done precisely where the series stops converging, and we find \(\abs{q} < 0.35(3)\dots\).
A precise estimate will be obtained in the next section.

Physical observables can be computed at any given order in the expansion parameters (either \(\tilde \mu\) or \(\Phi_0\) in the grand canonical picture, or \(q\) in the canonical one), but we did not find simple compact expressions.
For this reason we need to use numerical methods to study the analytic properties of the perturbative expansions.
This is done in the next section.

Even though we have the convergence under control it is still interesting to study the analytic structure of the series with other methods.
We will see that in the grandcanonical description we have complex poles, which disappear after the Legendre transform and there are only real (negative) poles in the expansion of the energy in the canonical ensemble.

\subsection{Analytic properties}
\label{sec:small-Q-N-analytic-properties}

\paragraph{Grandcanonical ensemble.}

\begin{figure}
\centering
\begin{tabular}{lr}
\includegraphics[width=0.45\textwidth]{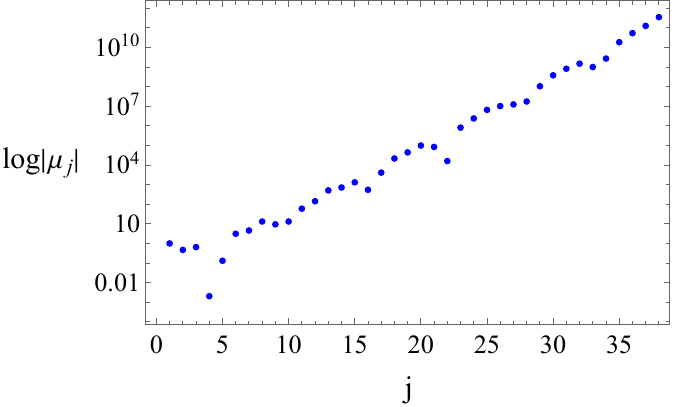} & 
\includegraphics[width=0.45\textwidth]{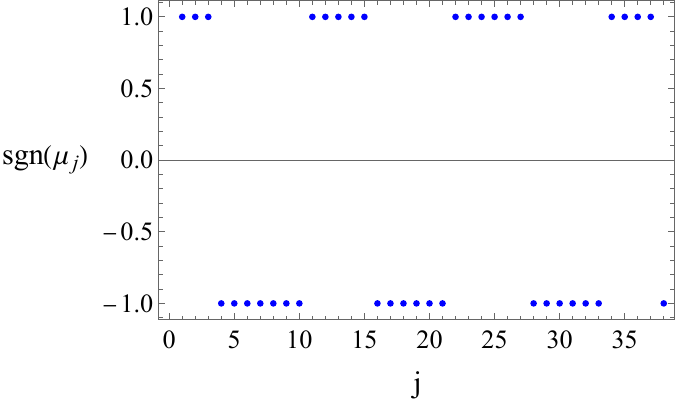}
\end{tabular}
	\caption{Absolute value (\emph{left}) and sign (\emph{right}) of the coefficients $\mu_j$ in Eq.~\eqref{gapseries} describing the chemical potential as a series in even powers of the gap.}
	\label{plotaj}
\end{figure}

It is interesting to determine the analytic properties of the perturbative series appearing in the small $q$ limit. 
Let us start from the grandcanonical picture, where the chemical potential is kept fixed.
As illustrated in  Fig.~\ref{plotaj}, the sign of the $\mu_j$ coefficients in Eq.~\eqref{gapseries}, describing the relationship between gap and chemical potential, fluctuates wildly and their magnitude exhibits an oscillatory behavior.
This is a typical sign of the presence of competing singularities in the complex $\Phi_0$ plane.

\begin{figure}
\centering
\includegraphics[ width=0.9\textwidth]{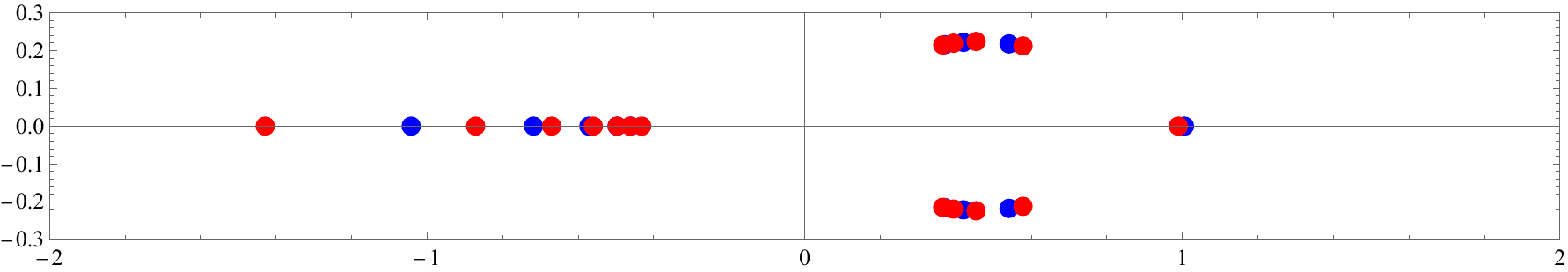} \\ \vspace{2cm}
\includegraphics[ width=0.9\textwidth]{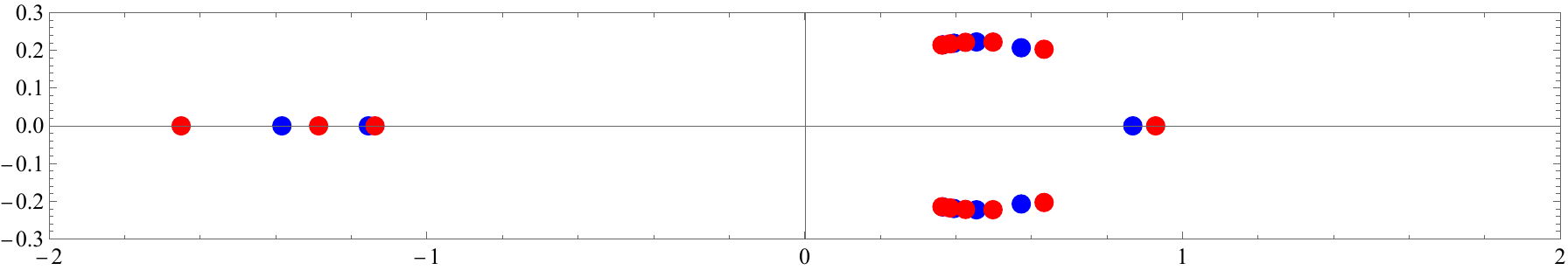}
	\caption{Singularities of the $(14,15)$ (blue) and $(20,18)$ (red) Padé approximants for the chemical potential $r \tilde \mu = \sum_{j=1}^{\infty} \mu_j (r \Phi_0)^{2j}$ (\emph{top}) and the grand potential $\frac{\Omega }{2 N}= \sum_{j=1}^{\infty} \Omega_j (r \Phi_0)^{2(j+1)}$ (\emph{bottom}), expressed as series in powers of the gap.}
	\label{pade}
\end{figure}

To probe the singularity structure, we study the distribution of poles in the Padé approximants.
As can be seen in Fig.~\ref{pade}~(top), the Padé approximants show an accumulation of poles on the negative real axis around $ r \Phi_0 \sim 0.45$ (in agreement with our previous estimate) as well as poles hinting at two complex conjugate singularities whose distance from the origin is approximately the same as the one on the real axis. 

This estimate of the radius of convergence is confirmed by the root test, amounting to studying the sequence $\abs{\mu_j}^{-1/j}$ which converges to the radius of convergence in the $j \to \infty$ limit. 

We find the same radius of convergence for the small $r \Phi_0$ expansion of the grand potential in Eq.~\eqref{potseries}.
However, in this case, the singularity on the negative real axis occurs at larger values of $ r \Phi_0$ and does not determine the radius of convergence, as can be seen from the pole structure of the Padé approximants in Fig.~\ref{pade}~(bottom).

\paragraph{Canonical ensemble.}

The analysis becomes more manageable in the canonical picture, where the charge is kept fixed.
When $\Phi_0$, $\tilde \mu$, and $\Delta$ are expressed as a series in the physical parameter $q$, the analytic structure of the small $q$ expansion simplifies considerably. 
In fact, the coefficients $\phi_j$ and $\delta_j$ appearing in Eq.~\eqref{gapserQ} and Eq.~\eqref{altj} show no oscillation in magnitude, and a simple sign alternating pattern.
This indicates that the singularity determining the radius of convergence lies on the negative real axis. 

\begin{figure}
  \centering
  \begin{tabular}{lr}
    \includegraphics[ width=0.45\textwidth]{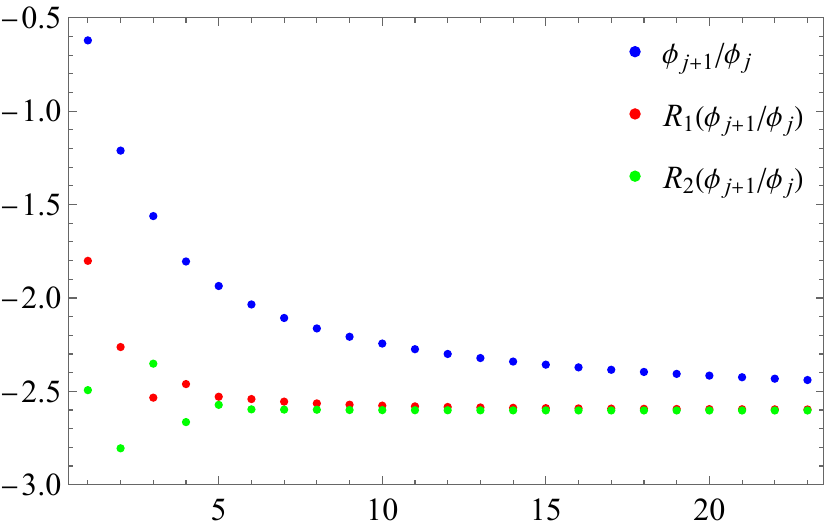} & \includegraphics[ width=0.45\textwidth]{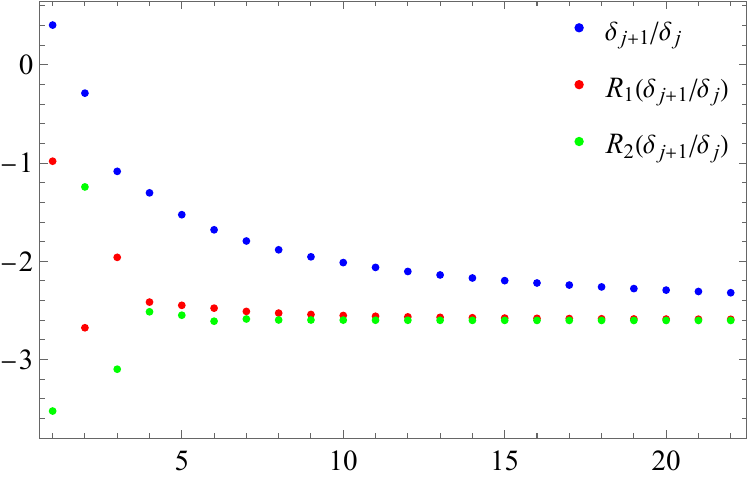}
  \end{tabular}
	\caption{Ratio of consecutive coefficients for the expansion of the gap~\eqref{gapserQ} (\emph{left}) and the scaling dimension~\eqref{altj} (\emph{right}) along with the first two Richardson transforms.}
	\label{ratio}
\end{figure}

To characterize the leading singularities we can use Darboux's theorem.
The singularity structure of a given function $\Psi(x)=\sum \psi_n x^n$ is fixed by the large-order behavior of the coefficients $\psi_n$ (see e.g.~\cite{Darboux,Dorigoni:2015dha}).
Concretely, if the $\psi_n$ satisfy
\begin{equation}
    \psi_n \underset{n \to \infty}{\sim} \frac{1}{x_0^n} \left[\psi(x_0) \binom{n+p-1}{n} - x_0 \psi'(x_0)  \binom{n+p-2}{n}+ \frac{x_0^2}{2!} \psi''(x_0) \binom{n+p-3}{n} -\dots \right] + \dots
    \label{darbo} \, ,
\end{equation}
then $x = x_0$ is the singularity determining the radius of convergence of the series expansion and, in the vicinity of $x_0$, the function behaves as
\begin{align}
	\label{leadsinger}
  \Psi(x)&= \psi(x) \left(1-\frac{x}{x_0} \right)^{-p} + \text{analytic} \,, & x \to & x_0 \,,
\end{align}
with $\psi(x)$ analytic near $x = x_0$.
The parameters in Eq.~\eqref{darbo} are determined by considering the $n \to \infty$ limit of certain sequences which can be constructed starting from the $\psi_n$.
For instance, it is well known that the radius of convergence $x_0$ can be determined by analyzing the ratio of consecutive coefficients $\psi_n/\psi_{n-1} \underset{n \to \infty}{\sim} 1/x_0 $.
Analogously, $p$ and $\psi(x_0)$ can be obtained by studying
\begin{align} \label{Eqforp}
    p &= 1 + \lim_{n \to \infty} n \left(x_0 \frac{\psi_n}{\psi_{n-1}} - 1 \right) \,, &
    \psi(x_0) &=\lim_{n \to \infty} \frac{\psi_n}{\left(\frac{1}{x_0}\right)^n \binom{n+p-1}{n}} \,.
\end{align}
The result of the ratio test for the coefficients in Eq.~\eqref{gapserQ} and~\eqref{altj} is illustrated in Fig.~\ref{ratio} and indicates a radius of convergence $q \sim  -0.384 \dots$, consistently with the results in the previous section.

Moreover, we found that in the vicinity of this point, the gap and the scaling dimension satisfy Eq.~\eqref{leadsinger} with $p=-1/2$, $\psi(x_0)\sim 0.253 \dots$ and $p=-3/2$, $\psi(x_0)\sim -0.060 \dots$, respectively.   The exponent is the same that has appeared in the bosonic case~\cite{Dondi:2021buw}.

In bosonic models, the radius of convergence admits a simple physical interpretation as the value for which the radial mode becomes gapless~\cite{Dondi:2021buw,Antipin:2022dsm}. While it is natural to expect the same to hold for the \ac{njl} model, the calculation of the mass of the radial mode on $\mathbb{R} \times S^2$ is not straightforward and goes beyond the scope of the present work.

\section{Asymptotic expansion for large Q/N}
\label{sec:largeQ}

We now move to the large $q$ expansion.
We know already that in the grandcanonical picture this limit is the same as large \(r \mu\) and \(r \Phi_0\).
In this regime, the behavior of the zeta function is well captured by a representation in terms of the Mellin transform of an appropriate heat kernel.
\begin{equation}
  \Tr*( \Sigma^{-s}) = \frac{1}{\Gamma(s)} \int_0^{\infty} \frac{\dd{t}}{t} t^s \Tr*( e^{-\Sigma t}) \, .
\end{equation}
Studying this heat kernel we will find a perturbative expansion in \(1/q\), which is now asymptotic and needs to be supplemented by exponential corrections.

\bigskip

Let us start by rewriting our zeta function as a Mellin transform:
\begin{equation} \label{Mellano}
  \begin{aligned}
    \zeta(s | \slashed{\del{}}, \mu, \Phi_0) &=2 r^{2s} \sum _{\ell=1}^{\infty } \ell \bqty*{ \pqty*{ \pqty*{\ell + r \mu}^2 + (r \Phi_0)^2 }^{-s} + \pqty*{ \pqty*{\ell - r \mu}^2 + (r\Phi_0)^2 }^{-s} }  \\
                                                      &=  \frac{2 }{ \Gamma (s)} \int_0^{\infty } \frac{\dd{t}}{t} t^s e^{-\Phi_0 ^2 t} \sum _{\ell=1}^{\infty } \ell \pqty*{ e^{-(\ell/r+\mu )^2 t}+e^{-(\ell/r-\mu )^2 t}} \, .
  \end{aligned}                                                        
\end{equation}
We are interested in the limit of large gap \(\Phi_0\) (in units of the inverse radius \(1/r\)).
Because of the exponential \(\exp[-\Phi_0^2 t]\), the integral localizes around the \(t \to 0\) limit of the heat kernel
\begin{equation}
  K(t|\mu) = \Tr*( e^{- t \Sigma }) = \sum _{\ell=1}^{\infty } \ell \pqty*{ e^{-(\ell/r+\mu )^2 t}+e^{-(\ell/r-\mu )^2 t}} \, ,
\end{equation}
where
\begin{equation}
  \Sigma = \pqty*{ \Gamma^{\mu} \nabla_{\mu} - \Gamma_3 \Gamma_5 \mu}^{\dagger} \pqty*{ \Gamma^{\mu} \nabla_{\mu} - \Gamma_3 \Gamma_5 \mu} .
\end{equation}

This form, however, is not adapted to taking the \(t \to 0\) limit because all the exponentials contribute at the same order.
To overcome this difficulty, we perform a Poisson resummation, \emph{i.e.} we use the identity
\begin{equation}
  \sum_{l \in \setZ} f(l) = \sum_{k \in \setZ} \int \dd{\rho} f(\rho) e^{2 \pi i k \rho}.
\end{equation}
The Fourier transforms can be evaluated analytically and we find
\begin{equation} \label{Heat}
  \begin{aligned}
    K(t | \mu) &= \sum _{\ell=1}^{\infty } \ell \left(e^{-(\ell/r+ \mu )^2 t} + e^{-(\ell/r-\mu )^2 t}\right) = \frac{1}{2} \sum _{\ell=-\infty }^{\infty } \abs{\ell} \left(e^{-(\ell/r+\mu )^2 t}+e^{-(\ell/r -\mu)^2 t}\right)  \\
    &
      \!\begin{multlined}
        =\frac{1}{2}\int_{-\infty }^{\infty }\dd{\rho} \abs{\rho} \left(e^{-(\mu +\rho/r )^2 t}+e^{-(\mu -\rho/r )^2 t}\right) \\
        + \frac{1}{2} \sideset{}{'}\sum _{k} \int_{-\infty }^{\infty } \dd{\rho} \abs{\rho}  \left(e^{-(\mu +\rho/r )^2 t-i 2 \pi  k \rho }+e^{-(\mu -\rho/r )^2 t -i 2 \pi  k \rho }\right) 
      \end{multlined}
    \\
    &
      \!\begin{multlined}
        = \frac{r^2}{t}\left(e^{- \mu ^2 t} + \mu \sqrt{\pi t }   \erf(\mu  \sqrt{t}) \right) \\
        + \frac{ r^2 e^{-\mu ^2 t}}{t^{3/2}} \sideset{}{'}\sum_{k} \left(\sqrt{t} - \left(\pi r  k+i \mu  t \right) F\mleft(\frac{\pi r k+i\mu t}{\sqrt{t}}\mright) - \left( \pi r k - i \mu  t \right) F\mleft(\frac{ \pi r k -i \mu t}{\sqrt{t}}\mright)\right),
      \end{multlined}
  \end{aligned}
\end{equation}
where the primed sum is over \(k \in \setZ \setminus \set{0}\), \(\erf(z)\) is the error function, and \(F(z)\) is Dawson's function,
\begin{equation}
  F(z) = \frac{\sqrt{\pi}}{2} e^{z^2} \erf(z) =  e^{-z^2} \int_0^z \dd{u} e^{-u^2} \ .
\end{equation}

The expression that we have obtained is suggestive of a sum over instantonic contributions.
To make this observation precise, we need to decompose Dawson's function into its perturbative and non-perturbative parts.

In the limit of interest, the absolute value of the argument of Dawson's function is large,
\begin{equation}
  \abs*{ \frac{\pi r k \pm i \mu t}{\sqrt{t}} }^2 = \frac{\pi^2 r^2 k^2}{t} + \mu^2 t \to \infty,
\end{equation}
and --- as we will see in the following --- the Mellin integral in the zeta function has a saddle at \(\Phi_0 t = k \pi r\).
It follows that the phase of the argument satisfies \(\tan(\theta) = \pm 1/\sqrt{\kappa_0^2 - 1} \), for an angle  \(\theta \approx \pm \pi/3 \), and we can consistently use the asymptotic expansion
\begin{equation}
  \label{eq:Dawson-expansion}
  F(x) \underset{x \to \infty}{=} \sum_{n=0} \frac{(2n-1)!!}{2^{n+1}}\pqty*{\frac{1}{x}}^{2n+1} \mp \frac{i\sqrt{\pi}}{2} e^{-x^2}.
\end{equation}
We have obtained a transseries representation~\cite{Ecalle,Dorigoni:2014hea,Aniceto:2018bis} of the heat kernel around \(t \to 0\).
The reality condition for \(F(x)\) is sufficient to fix the Stokes parameter in front of the exponential.
In turn this fixes all the ambiguities in the asymptotic expansion of the heat kernel \(K(t| \mu)\).

The perturbative and non perturbative parts in the expansion of the Dawson's function will turn directly into the large charge asymptotic expansion and exponential contributions.

\paragraph{Perturbative expansion.}
The perturbative part is written by using the binomial expansion for the argument of \(F\) and recognizing that the sum over \(k\) reduces to a Riemann zeta function.
After some manipulations, the perturbative part of the heat kernel reads
\begin{equation}
  \begin{aligned}
    K(t | \mu) &\sim %
    \sqrt{\frac{\pi}{t}} \mu r^2  \erf(\mu  \sqrt{t}) + e^{-\mu^2 t} r^2 \bqty*{\frac{1}{t} - \frac{1}{t}\sum_{m=1}^\infty \sum_{l=0}^{m-1} \frac{(-4)^l \abs{B_{2m}}}{m \Gamma(m - l) \Gamma(2l+1)} (\mu^2 t)^{l} \frac{t^m}{r^{2m}}} \\
    & \sim \sqrt{\frac{\pi}{t}} \mu r^2 \erf(\mu  \sqrt{t}) + e^{-\mu^2 t} \bqty*{\frac{r^2}{t} - \frac{1}{6} - \frac{t}{60 r^2} + \frac{t^2 \mu^2}{30 r^4} + \dots },
  \end{aligned}
\end{equation}
where the symbol \(\sim\) indicates asymptotic equivalence.
One advantage of having an explicit form for the series is that we can evaluate its large-order behavior.
From this form, we can see that the Mellin integral localizes around $t \sim \order{r/\mu}$.

{For this reason, we rewrite the double sum as a series expansion around $t=0$ at fixed values of the product $\mu t/r$ as 
\begin{equation} \label{Bor}
\sum_{m=1}^\infty \sum_{l=0}^{m-1} \frac{(-4)^l \abs{B_{2m}}}{m \Gamma(m - l) \Gamma(2l+1)} (\mu^2 t)^{l} \frac{t^m}{r^{2m}} =  \sum _{n=1}^{\infty } c_n   \frac{t^n} {r^{2n}},
\end{equation} 
where
\begin{equation} 
c_n = \sum _{l=0}^{\infty } \frac{ (-1)^{n+1} 4^l B_{2 (l+n)}}{(2 l)! (l+n) \Gamma (n)} \left(\frac{\mu t}{r}\right)^{2 l}.
\end{equation}
The sum over $l$ defining the $c_n$ coefficients has a finite radius of convergence stemming from the presence of two complex conjugate poles at $\mu t/r  = \pm i \pi $, where the argument of the Dawson function in Eq.~\eqref{Heat} vanishes.
The large-order behavior of the $c_n$ coefficients is\footnote{To find the large order behavior, one can analytically continue $\mu t/r$ to purely imaginary values and identify the value of $l$ which maximizes the coefficients.}
\begin{equation}
c_n \sim \frac{\Gamma (n)}{\sqrt{\pi}}   \left((\pi -i \mu t/r)^{1-2 n} + (\pi + i \mu t/r)^{1-2 n} \right) \,,
\end{equation}
and, therefore, the coefficients of the perturbative expansion diverge as $n!$. The large-order behavior of the coefficients implies that the Borel transform $\mathcal{B}\left[ \sum _{n=1}^{\infty } c_n   \frac{t^n} {r^{2n}} \right]$(z) of the perturbative expansion~\eqref{Bor} has a finite radius of convergence determined by two complex conjugate poles at 
\begin{equation}
z^* = (\pi \pm i \mu t/r)^2 \,.
\end{equation}
The convolution with the Borel transform of $e^{-\mu^2 t}$ (\emph{i.e.} a delta function centered at $(\mu t/r)^2$) shifts the position of the leading singularities to  $z^* = \pi^2 \pm 2 \pi i \mu t/r$. 
As we shall see below, these singularities are associated with the leading non-perturbative corrections to the perturbative expansion of the heat kernel. Moreover, in Appendix~\ref{app2} we will study the transseries representation of the heat kernel at fixed $\mu$ and show how, in this case, the nonperturbative corrections admit a simple geometric interpretation as worldline instantons.}

The Mellin transform of the heat kernel is
\begin{equation}
  \begin{aligned}
    \zeta(s | \slashed{\del{}}, \mu, \Phi_0) & \!\begin{multlined}[t] \sim \left( 2 \frac{\mu^2}{\Phi_0^{2s}} {\textstyle \pFq{2}{1}{1/2,s}{3/2}{-\frac{\mu^2}{\Phi_0^2}}} + \frac{1}{s-1} \pqty*{ \mu^2 + \Phi_0^2}^{1-s} \right) r^{2-s} \\
      - \sum_{m=1}^{\infty} \frac{\abs{B_{2m}}}{m} \pqty*{\sum_{l=0}^{m-1} \binom{m + l + s - 2}{2l, \mspace{6mu} m - l - 1, \mspace{6mu} s -1} (-4)^l \pqty*{1 + \frac{\Phi_0^2}{\mu^2}}^{-l}} \frac{\pqty*{\mu^2 + \Phi_0^2}^{1-s-m}}{r^{s-2+2m}}
    \end{multlined} \\
                                    & \!\begin{multlined}
                                      \sim \left( 2 \frac{\mu^2}{\Phi_0^{2s}} {\textstyle \pFq{2}{1}{1/2,s}{3/2}{-\frac{\mu^2}{\Phi_0^2}}} + \frac{1}{s-1} \pqty*{ \mu^2 + \Phi_0^2}^{1-s} \right) r^{2-s}  + \frac{1}{6} \pqty*{ \mu^2 + \Phi_0^2}^{-s} r^{-s}\\
                                      + \frac{s}{60} \pqty*{1 - \frac{2(1+s)\mu^2}{\mu^2 + \Phi_0^2}} \pqty*{ \mu^2 + \Phi_0^2}^{-s-1} r^{-s-2} + \dots,
                                          \end{multlined} 
  \end{aligned}
\end{equation}
where \(\pFq{2}{1}{a,b}{c}{z}\) is the Gauss hypergeometric function
\begin{equation}
  {\textstyle \pFq{2}{1}{a, b}{c}{z}} = \sum_{n=0}^{\infty} \frac{(a)_n (b)_n}{(c)_n} \frac{z^n}{n!} \ .
\end{equation}

As expected, we find an asymptotic series in \(\pqty*{\mu^2 + \Phi_0^2}^{-1}\) which, after minimization over the gap \(\Phi_0\) and a Legendre transform, turns into a series in \(1/Q\).
In this expression it is manifest that the term of order \(\order*{\pqty*{\mu^2 + \Phi_0^2}^{1-s-m}}\) is written as the sum of precisely \(m \) terms of order \(\order{\Phi_0/\mu} = \order{1}\).

The Mellin transform increases the order of divergence, and now the coefficients in the expansions grow for \(m \to \infty\) as \(\order{(2m)!}\). 
Because of this, from the general theory of transseries, we expect the non-perturbative part to be controlled by exponentials of the type \(e^{- (\mu^2 + \Phi_0^2)^{1/2}} \sim e^{- \sqrt{q}}\), consistently with the general behavior predicted in~\cite{Dondi:2021buw}.
To verify this prediction, we need to take into account the exponentially-suppressed terms in the expansion of Dawson's function in Eq.~\eqref{eq:Dawson-expansion}.

Before moving to that, we can use the explicit expression for the zeta function to write our physical observables in the large-\(q\) limit.
For the action on the zero modes we have
\begin{equation}
  \begin{aligned}
    \frac{S(\mu, \Phi_0)}{\beta N} &
                \!\begin{multlined}[t]
                  = \pqty*{\frac{\mu^2 - 2 \Phi_0^2}{3} \sqrt{\mu^2 + \Phi_0^2} + \mu \Phi_0^2 \arcsinh*(\frac{\mu}{\Phi_0})} r^2 \\
                  + \frac{1}{2 \sqrt{\pi}} \sum_{m=1}^{\infty} \frac{\abs{B_{2m}}}{m} \pqty*{\sum_{l=0}^{m-1} \frac{(-4)^l \Gamma(l+m-3/2)}{\Gamma(2l+1) \Gamma(m-l)} \pqty*{1 + \frac{\Phi_0^2}{\mu^2}}^{-l}} \pqty*{\mu^2 + \Phi_0^2}^{3/2-m} r^{2-2m}
                \end{multlined}
    \\
    & \!
      \begin{multlined}
        = \pqty*{\frac{\mu^2 - 2 \Phi_0^2}{3} \sqrt{\mu^2 + \Phi_0^2} + \mu \Phi_0^2 \arcsinh*(\frac{\mu}{\Phi_0})}r^2 +\frac{1}{6} \sqrt{\mu ^2+\Phi_0 ^2}  \\
        + \frac{\Phi_0 ^2}{120 \left(\mu ^2+\Phi_0 ^2\right)^{3/2} r^2} + \frac{\Phi_0 ^4-4 \mu ^2 \Phi_0 ^2}{1008 \left(\mu ^2+\Phi_0 ^2\right)^{7/2} r^4}+\frac{8 \mu ^4 \Phi_0 ^2-12 \mu ^2 \Phi_0 ^4+\Phi_0 ^6}{1920 \left(\mu ^2+\Phi_0 ^2\right)^{11/2} r^6} + \dots
        \end{multlined}
  \end{aligned}
\end{equation}
At the saddle, \(\mu\) and \(\Phi_0\) are of the same order \(\mu r = \order{q^{1/2}}\).
It follows that the terms proportional to \(r^{n-1}\) will contribute to the energy proportionally to \(q^{n/2}\).

Explicitly, the solution to the gap equation \eqref{gappo} reads
\begin{align}
     \Phi_0 r &= \sqrt{\kappa_0^2 -1} \left(\mu r -\frac{1}{12 \kappa_0^2} \frac{1}{\mu r} + \frac{33-16 \kappa_0^2}{1440 \kappa_0^6}\frac{1}{(\mu r)^3} -\frac{1464 \kappa_0^4-9663 \kappa_0^2+9926}{362880 \kappa_0^{10}}\frac{1}{(\mu r)^5} + \dots \right) \,,
\end{align}
while the scaling dimension is given by
\begin{align}
   \frac{ \Delta}{2 N} = \frac23 \left(\frac{q}{\kappa_0}\right)^{3/2}+ \frac16 \left(\frac{q}{\kappa_0}\right)^{1/2} + \frac{11-6 \kappa_0^2}{720 \kappa_0^2}\left(\frac{q}{\kappa_0}\right)^{-1/2} + \dots \,.
\end{align}

\paragraph{Non-perturbative contributions.}

Let us now move on to the exponential part of the transseries expansion.
Using the expansion of Dawson's function in Eq.~\eqref{eq:Dawson-expansion} we find for the heat kernel
\begin{align} \label{noce}
	K(t | \mu) &\ni i\frac{\sqrt{\pi} r^2}{2 t^{3/2}} e^{-\mu ^2 t} \sideset{}{'}\sum_k  \left((\pi  k r+i \mu  t)  e^{-\frac{\left(\pi  k r+i \mu  t\right)^2}{t}} +  (\pi  k r-i \mu  t) e^{-\frac{\left(\pi  k r-i \mu  t\right)^2}{t}} \right) \nonumber \\ &
    =i\frac{\sqrt{\pi} r^2}{t^{3/2}} \sideset{}{'}\sum_k\sign(k)\pqty*{k \pi r \cos(2\pi k \mu r) + t \mu \sin(2\pi k\mu r)}e^{-k^2 \pi^2 r^2/t}.
\end{align}
In the exponents appearing on the first line, we recognize the position of the leading singularity (corresponding to $k=1$) of the Borel transform of the heat kernel expansion in Eq.~\eqref{Bor}.
Correspondingly, the nonperturbative part of the expression for the zeta function reads
\begin{multline}
	 \label{transer}
   \zeta(s | \slashed{\del{}}, \mu, \Phi_0) 
	\ni  
	\frac{r^{2}}{\Gamma(s)} \int_0^\infty \frac{\dd{t}}{t} t^s e^{-\Phi_0^2 t} \frac{i\sqrt{\pi}}{t^{3/2}} \sideset{}{'}\sum_k \sign(k) \Big( k \pi r \cos(2\pi k \mu r) \\
	+t\mu \sin(2\pi k \mu r)\Big) e^{-k^2\pi^2 r^2/t}.
\end{multline}
The exponential is suggestive of the action for a particle of mass $\Phi_0$ propagating on geodesics on the two-sphere winding \(k\) times.
To see this immediately, we can simply approximate the integral with its saddle point by writing
\begin{equation}
	\exp*[-\Phi_0^2 t - \frac{k^2\pi^2 r^2}{t}] = \exp*[- \pqty*{\Phi_0\sqrt{t} - \frac{k \pi r}{\sqrt{t}}}^2 - 2 \pi k r \Phi_0],
\end{equation}
and get directly the expected exponential contribution $e^{-2\pi \abs{k} r \Phi_0}$ at the saddle where \(t = k \pi r/\Phi_0\).

This estimate can be made precise by solving the integrals explicitly in terms of Bessel functions.
The integral can be solved explicitly:
\begin{multline}
  \zeta(s | \slashed{\del{}}, \mu, \Phi_0) \ni \frac{2i\pi^s r^{s+3/2}}{\Gamma(s)}\pqty*{\frac{\Phi_0}{\abs{k}}}^{1/2-s} \big[\Phi_0 \cos(2\pi k\mu r )K_{3/2-s}(2\pi r \Phi_0\abs{k} r) \\
  + \mu \sin(2\pi k \mu r )\sign(k) K_{1/2-s}(2\pi r \Phi_0\abs{k} r) \big] \ .
\end{multline}
Setting $s= -1/2$ we find the contribution to the action for the zero modes:
\begin{equation}
\frac{S(\mu, \Phi_0)}{\beta N} \ni	\frac{i\Phi_0 r}{k\pi}\pqty*{{\Phi_0}\sign(k)\cos(2\pi k\mu r)K_{2}(2\pi \Phi_0\abs{k} r) + \mu \sin(2\pi k\mu r) K_{1}(2\pi \Phi_0\abs{k} r)},
\end{equation}
which we can expand for large $\Phi_0$ to recover the expected exponential terms:
\begin{multline}
	\frac{S(\mu, \Phi_0)}{\beta N} \ni	-\frac{i}{\abs{k}^{3/2}32\pi^2\sqrt{\Phi_0 r}} e^{-2\pi \Phi_0\abs{k} r} \Bigg[ \abs{k} \Phi_0  \cos(2\pi k \mu r) \pqty*{ 16\pi \Phi_0 r+\frac{15}{\abs{k}}+\dots }  \\
	+ \sign(k) \mu \sin(2\pi k \mu r)(16 \pi \Phi_0\abs{k} r+ 3 +\dots )\Bigg].
\end{multline}

From this expression one can immediately derive the grand potential \(\Omega\) using the gap equation.
Using the exact form for \(S(\mu, \Phi_0)\) we can write the observables at any given order in the charge.
For illustrative purposes, we will limit ourselves here to the leading terms in the perturbative expansion associated to each exponential term.
Using
\begin{equation}
	\Phi_0 = \mu  \sqrt{\kappa_0^2 - 1} \, , 
\end{equation}
with \(\kappa_0 \tanh(\kappa_0) = 1\), we find for the grand potential
\begin{equation}
 \frac{ \Omega}{2 N} \ni - i \frac{ \pqty*{\kappa_0^2 - 1}^{1/4} r^{1/2}\mu^{3/2}  e^{-2 \pi r \sqrt{\kappa_0^2 - 1} \abs{k} \mu} }{2 \pi \sqrt{\abs{k}}}   \pqty*{ \sqrt{\kappa_0^2 - 1} \pqty*{\cos(2 \pi k \mu r) + \dots } + \pqty*{\sin(2 \pi k \mu r) + \dots } }
\end{equation}
and for the conformal dimension of the lowest operator of charge \(Q\):
\begin{equation}
  \frac{\Delta(Q)}{2N} \ni - i \frac{ \pqty*{\kappa_0^2 - 1}^{1/4} q^{3/4} e^{-2 \pi  \sqrt{\kappa_0^{-1} - \kappa_0^{-3}}\abs{k} \sqrt{q}}}{2 \pi \kappa_0^{9/4} \sqrt{\abs{k}}}  \pqty*{ \sqrt{\kappa_0^2 - 1} \cos*(\frac{2 \pi k}{\kappa_0^{3/2}} \sqrt{q}) +  \sin*(\frac{2 \pi k}{\kappa_0^{3/2}} \sqrt{q})    }  +  \dots
\end{equation}
This final result is consistent with the prediction of~\cite{Dondi:2021buw}. The nonperturbative effects are controlled by $e^{-\alpha \sqrt{q}}$, indicating that the perturbative series is asymptotic and diverges like $(2n)!$. It also follows that the optimal truncation is obtained at $\order{\sqrt{q}}$ terms.
This qualitative behavior seems to be general for gapped systems at fixed charge. 
The geometric interpretation via worldline instantons provides a robust unifying framework for these systems.

\section{Geometric interpretation: worldline instantons}
\label{sec:geomInterp}

The resurgence analysis of the previous section has led us to a transseries expansion for the one-loop determinant, closely resembling that found for the bosonic $O(N)$ model in the large $N$-limit~\cite{Dondi:2021buw}. In that case, the exponential corrections admitted a natural interpretation in terms of worldline instantons.

The worldline formalism has proven to be an effective first-quantized alternative for determining quantum field theory amplitudes, anomalies, and effective actions in the presence of background fields~\cite{Schubert:2001he,Bastianelli:2006rx}.
Within this framework, functional determinants are expressed in terms of quantum-mechanical path integrals on the manifold of interest.
In this way, one obtains an independent non-perturbative definition of the physical observables, which resolves the ambiguities (Stokes parameters) that generically appear in the resurgence analysis.\footnote{In our case, the ambiguities are solved by requiring the effective action to be real.} 
In particular, in the bosonic case studied in~\cite{Dondi:2021buw}, the quantum-mechanical path integral associated with the heat kernel can be evaluated via a saddle-point expansion in the $t \to 0$ limit. Consequently, geodesics winding $k$ times around the sphere produce contributions of the form $e^{-S_I/t}$ with instantonic action $S_I =(\pi k r)^2$.
In the same paper it was also conjectured that this geometrical interpretation is robust and can be extended to describe the non-perturbative behavior of the large-charge expansion of general systems.

In the case at hand, the exponential corrections to the \ac{njl} heat kernel in Eq.~\eqref{noce} share indeed the same structure, which is suggestive of a similar geometric interpretation.
Different formalisms have been proposed in the literature for the worldline description of the one-loop effective action for a Dirac particle coupled to an axial gauge field $a_{5\nu} = \delta_{\nu 0} \mu $~\cite{Strassler:1992zr,DHoker:1995aat,DHoker:1995uyv,McKeon:1998et,Bastianelli:2002qw,Bastianelli:2024vkp}.
Adopting the proposal developed in~\cite{McKeon:1998et,Bastianelli:2024vkp}, in App.~\ref{app2} we demonstrate that the transseries expansion of the heat kernel around $t=0$ at fixed $\mu$ can again be understood in terms of worldline instantons with action $S_I$. In this regime, the term $k \pi r \cos(2\pi k \mu r)$ in Eq.~\eqref{noce} arises from the determinant of the fluctuations around the winding geodesic while the $t \mu \sin(2\pi k\mu r)$ term represents a two-loop correction which we do not reproduce here.

However, as explained in Section~\ref{sec:largeQ}, in the case at hand, the Mellin integral localizes around $t \sim r/\mu $ so $t$ does not act as an independent perturbative parameter. Accordingly, the $t \to 0$ limit has to be considered at fixed $\mu t/r$. In this regime, the terms $\cos(2\pi k \mu r)$ and $\sin(2\pi k \mu r)$ in Eq.~\eqref{noce} appear to be of the same order as the saddle point action. 
This suggests that they may originate from the exponentiation of loop corrections around the instanton background or that one has to consider more involved saddle points forming a complex conjugate pair. A similar situation was encountered in~\cite{Demulder:2016mja,Dorigoni:2015dha}, where classical configurations related to complex poles in the Borel plane were found by considering complexified fields in the \ac{eft}. Needless to say, such scenarios considerably complicate the worldline analysis. Furthermore, as previously noted, there are differing proposals for how to formulate the worldline path integral, which are not obviously equivalent. We intend to return to this issue in future work.

\section{Conclusions}
\label{sec:conclusions}

The large-charge approach is a powerful approach for studying strongly coupled physics in any model with a global symmetry and allows writing a perturbative expansion in inverse powers of the charge in the \ac{eft} on the sector of fixed charge. It works best at a fixed point of the renormalization group due to the advantages of conformality. When combined with a second controlling parameter such as in a large-$N$ or small $\varepsilon$ expansion, it becomes possible to go beyond the perturbative sector and determine the exponential corrections. 

In this article, we have studied the three-dimensional \ac{njl} model in a double-scaling limit where both the charge $Q$ and the number of fermion flavors $N$ become large with a fixed ratio. We have followed here on the footsteps of~\cite{Alvarez-Gaume:2019biu,Dondi:2021buw}, where a similar analysis has been performed for the bosonic case of the O(N) model. In the bosonic case, the system is in a superfluid phase at large charge and the \ac{eft} is described by the Goldstone bosons of the broken symmetry. Fermionic models pose new challenges. We perform a Stratonovich transform in the large-$N$ limit at fixed charge and find that the condensate in the large-charge sector is due to Cooper pairs - the collective field contains two fermions and is charged under the global symmetry. The \ac{njl} model was studied both in the limit of large and small $Q/N$ in~\cite{Dondi:2022zna}. In the present work, we have studied these limits systematically beyond the first few orders and performed the resurgence analysis, identifying the exponential corrections. We also proposed a geometric interpretation in terms of the worldline of a particle moving on geodesics on the cylinder. We found that the conjecture stated in~\cite{Dondi:2021buw} that the nonperturbative corrections have the form $e^{\alpha \sqrt{Q}}$ holds true also for the \ac{njl} model.

Our results strongly suggest a worldline interpretation, but we lack a detailed construction of the action and the saddle. In the small-charge limit, we have seen that the structure of the singularities in the Borel plane greatly simplifies in the canonical ensemble. This observation might provide a clue towards the correct worldline action, as a similar simplification may also occur at large charge.

In this work, we have concentrated on the \ac{njl} model which has the crucial advantage of developing a condensate perturbatively in $1/N$. This is not the case for the Gross--Neveu model~\cite{Dondi:2022zna}. Instead, the large-$N$ behavior is the one of a Fermi surface, which must nonetheless display a \ac{bcs} instability. It would be interesting to understand how to adapt our analysis to this case.

\newpage

 \section*{Acknowledgments}
 {\dosserif
   We thank Daniele Dorigoni for illuminating discussions and comments on the draft.

 The work of J.B. and S.H was supported by the World Premier International Research Center Initiative (WPI Initiative), MEXT, Japan. The work of J.B. was also supported by the JSPS KAKENHI Grant Number JP23K19047. The work of S.R. was supported by the Swiss National
 Science Foundation under grant number 200021\_219267. This research was supported in part by grant NSF PHY-1748958 to the Kavli Institute for Theoretical Physics (KITP). S.H, D.O., and S.R. gratefully acknowledges support from the Simons Center for Geometry and Physics, Stony Brook University at which some of the research for this paper was performed. D.O. and S.R. acknowledge the support and hospitality of the Yukawa Institute for Theoretical Physics, Kyoto University, where part of this research was performed.
 }

\appendix

\section{Gap equation and grand potential via the Euler--Maclaurin summation formula} \label{Eulero}

In this appendix we provide alternative representations for the gap equation and the grand potential obtained by employing the Euler--Maclaurin summation formula.
The sums over $\ell$ are regularized by subtracting the \ac{uv} divergent terms and adding them back in the Zeta-function regularized form. For the gap equation, we obtain
\begin{align}
      \left(2 \tanh ^{-1}\left(\frac{1}{\sqrt{x^2+1}}\right)-2 \sqrt{x^2+1}\right)=\sum _{m=1}^{\infty } \frac{B_{2 m}}{m} {\textstyle \pFq{2}{1}{m-\frac{1}{2},m}{1}{-x^2}} (\mu r) ^{-2 m} ,
\end{align}
where $x \equiv \Phi_0/\mu$. This expression is of practical use for solving the gap equation in the large $q$ limit. Analogously, the grand potential $\omega$ reads
\begin{align}
    \frac{\omega}{r^2 \mu ^3}=  x^2 \tanh ^{-1}\left(\frac{1}{\sqrt{x^2+1}}\right) - \frac{1}{3} \left(1-2 x^2\right) \sqrt{x^2+1}+\sum _{m=1}^{\infty } \frac{B_{2 m}}{m} (\mu r)^{-2 m} \sum _{n=0}^{\infty } \binom{\frac{1}{2}}{n} \binom{1-2 n}{2 m-2} x^{2 n} \,.
\end{align}

\section{Transseries structure of the heat kernel at fixed chemical potential and geometric interpretation} \label{app2}

\subsection{Borel transform and exponential corrections}

We rewrite the perturbative expansion of the heat kernel in Eq.~\eqref{Heat} as a series around $t=0$ at fixed chemical potential $\mu$ as follows
\be
K(t | \mu) \sim    \sqrt{\frac{\pi}{t}} \mu r^2  \erf(\mu  \sqrt{t}) + e^{-\mu^2 t} r^2 \bqty*{\frac{1}{t} - \frac{1}{t} \sum_{n=1}^\infty a_n \left(\frac{t}{r^2}\right)^n } \,,
\ee
where 
\be
a_n =  \sum _{l=0}^{\left\lfloor \frac{n-1}{2}\right\rfloor } \frac{(-1)^n  4^l B_{2 n-2 l}}{(2 l)! (l-n) \Gamma (n-2 l)}\mu ^{2 l} \,.
\ee
Asymptotically, the coefficients behave as
\be
a_n \sim a(\mu) \ \pi ^{-2 n} \Gamma \left(n+\frac{1}{2}\right) \,,
\ee
where $a(\mu)$ is a function of $\mu$ whose exact form does not affect the location of the leading pole in the Borel plane, which occurs at $z= (\pi r)^2$ with $z$ denoting the Borel parameter. As expected, the position of the pole corresponds to the leading exponential correction, as evident from the second line of Eq.~\eqref{noce} evaluated for $k=1$. In contrast to the fixed $\mu t/r$ case considered in the main text, the sum defining the perturbative coefficients is convergent. However, the leading Borel-plane singularity now lies on the positive real axis, making the perturbative expansion non-Borel summable.

\subsection{Worldline representation}

In this section, we show that the exponential corrections in the transseries representation appearing in the heat kernel $K(t \rvert \mu)$ in the $t \to 0$ limit at fixed $\mu$ admit a simple geometric interpretation in terms of worldline instantons.

Specifically, we are interested in the determinant of the differential operator appearing in the \ac{njl} action~\eqref{HSaction}, which, on curved spacetime, reads
\begin{equation}
D = \Gamma^\mu \left(\nabla_\mu - \Gamma_5 a_{5 \mu} \right) +\Phi  P_+ + \bar{\Phi } P_-  \,, \qquad a_{5\nu} =\mu \delta_{\nu3} ,
\end{equation}
where $\Gamma^\mu = e_k^\mu \Gamma^k$ with $\Gamma^k$ the flat gamma matrices and $e_k^\mu$ the vielbein. The gamma matrices with curved indices are Hermitian and satisfy
\begin{equation}
\left \{ \Gamma_\mu, \Gamma_\nu\right \} =2 g_{\mu \nu} .
\end{equation}
The covariant derivative reads
\begin{equation}
\nabla_\mu  =\partial_\mu +\frac{1}{4} \omega_{\mu i j}  \Gamma^i \Gamma^j ,
\end{equation}
where 
\begin{equation}
 \omega_{\mu i j} = -\delta _{ik} e_j^{\nu } \left(\partial_\mu e_{\nu }^k-\Gamma _{\mu \nu }^{\rho } e_{\rho }^k\right) \,,
\end{equation}
is the spin connection. In order to derive the worldline representation of $\log \det D$, we consider the negative-definite operator
\begin{equation}
\Sigma \equiv D^\dagger D -\Phi_0^2 = -\Gamma ^{\mu }\Gamma ^{\nu }\nabla _{\mu }\nabla _{\nu } -2\Gamma _5 \left( \Gamma ^{\mu } \Gamma ^{\nu }-g^{\mu \nu } \right)a_{5 \nu }\nabla _{\mu }  +g^{\mu \nu }a_{5\mu } a_{5\nu }  ,
\end{equation}
where we replaced the Hubbard--Stratonovich field with its constant \ac{vev} $\Phi_0$. As usual, the Schrödinger--Lichnerowicz formula yields 
\begin{align}
\Gamma ^{\mu }\Gamma ^{\nu }\nabla _{\mu }\nabla _{\nu }  =  g^{\mu \nu } \nabla _{\mu }\nabla _{\nu }-\frac{R}{4} \ ,
\end{align}
where $R$ is the Ricci scalar of the manifold.
With this preparatory work, we can write Schwinger’s representation of $\log \det D$ as
\begin{align} \label{swing}
    \log \det D &=- \frac{1}{2} \int_0^\infty \frac{\dd{t}}{t}  e ^{- \Phi_0^2  t} \Tr e^{-\Sigma t} .       
\end{align}
Comparing the above to Eq.~\eqref{Mellano}, one can see that the functional trace is proportional to the heat kernel on the sphere once we take into account the contribution of the mode propagating around the thermal circle:
\begin{align} \label{sigmaris}
   \frac{1}{\beta}\Tr e^{-\Sigma t} = \frac{2}{\sqrt{\pi t}} K(t \rvert \mu) =\frac{2 r^2}{\sqrt{\pi } t^{3/2}}\left(1+\order{t} \right) \pm \frac{2  \pi i  r^3}{t^2} \sideset{}{'}\sum_k | k|  e^{-\frac{\pi ^2 k^2 r^2}{t}} \cos (2 \pi  k \mu  r) \left(1+\order{t} \right) \,.
\end{align}
In the worldline formalism, $\Sigma$ is interpreted as the Hamiltonian for a quantum mechanical particle with proper time $\tau$ that moves on a manifold $\mathcal{M}$ equipped with metric $g_{\mu \nu }$ in a given coordinate system $x^{\mu }$. The trace can then be expressed as a quantum mechanical path integral, given by~\cite{McKeon:1998et,Schubert:2001he,Bastianelli:2024vkp}

\begin{align}
    \Tr e^{-\Sigma t}    = \int \DD{x^{\mu}}
 \DD{\psi^\mu} \DD{\psi^4} e^{-S_{WL}[x^\mu,\psi^\mu, \psi^4,x^{\mu}; g_{\mu \nu}, a_{5 \mu} ]}   \,,
 \label{pathfinal}
\end{align}
with
\begin{align}
    S_{WL} =
\frac{1}{4 t} \int_{0}^{1} \! \! d\tau\, \Bigg [ &  \psi^4 \dot{\psi }^4 +
 g_{\mu \nu }\Big( \dot{x}^{\mu } \dot{x}^{\nu } + \psi ^{\mu } \dot{\psi }^{\nu }  +\psi ^{\mu } \left( \Gamma _{\lambda \rho }^{\nu }\dot{x}^{\lambda } - 4 \hat{\gamma}_5 g_{\rho \sigma }\dot{x}^{\nu } a_5^{\sigma }   \right) \psi ^{\rho }  \Big )\Bigg] \ .
\label{gfactionfinal}
\end{align}
Here $\psi^\mu(\tau)$ and $\psi^4(\tau)$ are Grassmann fields introduced to account for the trace over the Clifford algebra representation space. For the sake of simplicity, we have omitted all the terms that contribute beyond leading order in the semiclassical expansion around $t = 0$. Notably, these include the ghost fields as well as the counterterms arising from the nontrivial background geometry and the presence of the axial gauge field~\cite{Bastianelli:2006rx,Bastianelli:2024vkp} \footnote{However, these contributions cannot be neglected when considering the $t \to 0$ limit at fixed $\mu r/t$.}. 
The bosonic coordinates $x^\mu(\tau)$ obey periodic boundary conditions $x^\mu(0) = x^\mu(1)$, while the operator $\hat{\gamma}_5$, originating from the $\Gamma_5$ matrix, keeps track of the boundary conditions for the Grassmann fields. Specifically, each term in the expansion of the interaction exponential must be evaluated using antiperiodic (periodic) boundary conditions for  $\psi^\mu(\tau)$ and $\psi^4(\tau)$ depending on whether it contains $\hat{\gamma}_5$ raised to an even (odd) power. Equipped with the worldline path integral above, we now proceed to its semiclassical evaluation on $\mathbb{R}\times S^2$.

\subsection{Semiclassical calculation on the sphere}

We here show that the semiclassical expansion of the worldline path integral matches the transseries structure in Eq.~\eqref{noce}. 
To this end, we parameterize $S^2 \times \mathbbm{R}$ with $(\theta, \phi, s)$, \emph{i.e.} we use spherical coordinates on $S^2$. The metric reads
\be
g_{\mu \nu} =\text{diag}(r^2, r^2 \sin^2 \theta, 1) .
\ee
and the nonzero components of the Christoffel symbols are
\begin{align}
	\Gamma _{\theta \phi }^{\phi } &= \Gamma _{\phi \theta}^{\phi } = \cot \theta \,, & \Gamma _{\phi \phi }^{\theta } &= -\frac{1}{2} \sin (2 \theta ).
\end{align}
Therefore, the worldline action \eqref{gfactionfinal} reduces to
\begin{align}
       S_{WL}  = \frac{1 }{4 t}\int_0^1 d\tau &\Bigg[\psi^4 \dot{\psi }^4 +\psi ^s \dot{\psi }^s+\dot{s}^2 + r^2 \Bigg(\dot{\phi }^2 \sin ^2\theta +\dot{\theta }^2 + \psi ^{\theta } \dot{\psi }^{\theta }+\sin ^2\theta   \psi ^{\phi }\dot{\psi }^{\phi }  \nonumber \\ & +\frac{1}{2}  \sin (2 \theta ) \psi ^{\phi } \dot{\theta } \psi ^{\phi }- \sin (2 \theta )  \psi ^{\theta }\dot{\phi } \psi ^{\phi }-4 \mu   \psi ^{\theta } \dot{\theta }  \psi ^s-4 \mu   \sin ^2\theta   \psi ^{\phi }\dot{\phi }\psi ^s \Bigg)\Bigg] \, .
\end{align}
\paragraph{The classical solution to the \acs{eom}.}

We have periodic boundary conditions for the coordinates, 
\begin{align}
	\theta(1) &= \theta(0) = \pi/2 \,, & \phi(1) &= \phi(0) = 0,
\end{align}
and antiperiodic ones for the fermions. We consider a zero classical profile for the fermions and, therefore, the remaining \ac{eom} are the same as for the bosonic case~\cite{Dondi:2021buw},
\begin{align}
    2 \ddot\theta -\sin (2 \theta ) \dot\phi ^2&=0 \,, & \ddot\phi+  2 \dot\theta \dot \phi \cot \theta &=0,
\end{align}
with solution
\begin{align}
    \theta_0 &= \pi/2 \,, & \phi_0 &= 2\pi k \tau \, ,
\end{align}
which describes a particle winding \(k\) times around the great circle of the two-sphere.
The resulting saddle-point action is the same as in the bosonic case,
\be
S_I =  (\pi k r)^2 .
\ee
In this limit the dominating pole is on the real axis of the Borel plane and the trigonometric terms in the expression of the heat kernel are understood as coming from the quantum fluctuations that we compute below.

\paragraph{Quantum fluctuations.}
We denote the fluctuations as $h_\theta$, $h_\phi$, $h_s$ for the boson and $\chi_\theta$, $\chi_\phi$, $\chi_s$, $\chi_4$ for the fermions. The quadratic Lagrangian reads
\begin{align}
    \mathcal{L}_2 &=   \mathcal{L}_{bos}+\mathcal{L}_{fm},\\
    \mathcal{L}_{bos} &= \frac{r^2}{4 t} \Big[ \dot h_s^2 + \dot h_\phi^2 + \dot h_\theta^2-(2 \pi  k)^2 h_\theta^2  \Big],\\
    \mathcal{L}_{fm} &=  \frac{r^2}{4 t} \left[\chi_\theta  \chi_\theta +\chi_\phi  \dot  \chi_\phi -8 \pi  k \mu  \chi_s \chi_\phi +\chi_s \dot  \chi_s + \chi_4 \dot \chi_4 \right].
\end{align}
In what follows, we will take into account the phase factors related to the Morse index arising from the presence of negative modes by fixing the overall phase by imposing a reality condition on the heat kernel.
The bosonic determinant has already been calculated in~\cite{Dondi:2021buw} and reads \footnote{We adopt the standard worldline normalization~\cite{Schubert:2001he,Reuter:1996zm}. In particular, the free bosonic and fermionic path integrals are normalized to $\frac{1}{\sqrt{4 \pi t}}$ and unity, respectively.}. 
\begin{align}
\int_{h(1) = h (0) = 0} \DD{h}  e^{-\int_0^1  \mathcal{L}_{bos} d \tau} &
=    \frac{\pi | k| r^3}{2t^2} \,.
\end{align}
The fermionic fluctuations $\chi_\theta$ and $\chi_4$ are free and contribute only an overall normalization factor. By introducing $\eta = (\chi_s, \chi_\phi)$, the remaining functional determinant can be written as
\be \label{fiffi}
\int \DD{\eta} e^{-\frac{1}{2}\int_0^1 d \tau \eta \cdot \left[\frac{r^2}{2 t} \left(\partial_\tau - 2 i F \right)\right] \cdot \eta }  = \left(\det \left[ \frac{r^2}{2 t}  \partial_\tau\right]_A\right)^{1/2}\left(\det \left[\identity -2 i F \partial_\tau^{-1}\right]_A\right)^{1/2}  \,,
\ee
where  
\be
F= \left(
\begin{array}{cc}
 0 & -2 i \pi  k \mu  r \\
 2 i \pi  k \mu  r & 0 \\
\end{array}
\right).
\ee
The subscript $A$ indicates that the fermionic fluctuations are anti-periodic on the time circle with the associated spectrum of $\partial_\tau$ being $2\pi  i\left(n+1/2 \right)$. We, therefore, obtain 
\be
\left(\det \left[\identity -2 i F \partial_\tau^{-1}\right]_A\right)^{1/2} =\left( \prod _{n=-\infty }^{\infty } \left(1-\frac{4 k^2 \mu ^2 r^2}{(n+1/2)^2}\right)\right)^{1/2} = \cos(2 \pi k \mu r) \,.
\ee
Collecting all the results, we reproduce the expression in Eq.~\eqref{sigmaris}.

\printbibliography

\end{document}